\documentclass[twocolumn,aps, prl, showpacs, preprintnumbers, amsmath, amssymb, a4paper, superscriptaddress]{revtex4-2}

\usepackage{graphicx}
\usepackage{dcolumn}
\usepackage{bm}

\begin{document}

\preprint{APS/123-QED}

\title{Predicting magnetic edge behaviour in graphene using neural networks}

\author{Meriç E. Kucukbas}
\affiliation{School of Physics, Trinity College Dublin, Dublin 2, Ireland}
\author{Se\'an McCann}%
\affiliation{School of Physics, Trinity College Dublin, Dublin 2, Ireland}
\author{Stephen R. Power}
\email{stephen.r.power@dcu.ie}
\affiliation{School of Physics, Trinity College Dublin, Dublin 2, Ireland}
\affiliation{School of Physical Sciences, Dublin City University, Glasnevin, Dublin 9, Ireland}

\date{\today}

\begin{abstract}
  Magnetic moments near zigzag edges in graphene allow complex nanostructures with customised spin properties to be realised.
	However, computational costs restrict theoretical investigations to small or perfectly periodic structures.
	Here we demonstrate that a machine-learning approach, using only geometric input, can accurately estimate magnetic moment profiles, allowing arbitrarily large and disordered systems to be quickly simulated.
	Excellent agreement is found with mean-field Hubbard calculations, and important electronic, magnetic and transport properties are reproduced using the estimated profiles. 
	This approach allows the magnetic moments of experimental-scale systems to be quickly and accurately predicted, and will speed-up the identification of promising geometries for spintronic applications.
	While machine-learning approaches to many-body interactions have largely been limited to exact solutions of complex models at very small scales, this work establishes that they can be successfully applied at very large scales at mean-field levels of accuracy.
\end{abstract}

\maketitle


Graphene systems with zigzag (ZZ) edges, or mixed \emph{chiral} edges, can exhibit magnetic polarization at half-filling~\cite{Fujita1996, Son2006, Yazyev2010, Yazyev2011}.
A ZZ edge hosts a localised state at the Fermi energy~\cite{Nakada1996}, which is spin-split by electron-electron interactions\cite{Son2006, Son2006b}.
The resultant magnetic moments decay away from the edge and display an antiferromagnetic (AFM) texture with respect to sublattice, with different sublattice edges displaying opposite polarizations~\cite{Lieb1989, Fujita1996, Son2006, Palacios2008, Yu2008}.
Local moment formation and long spin-diffusion lengths together suggest graphene as a promising spintronic material~\cite{Han2014}, and many device proposals are predicated on ZZ-edge magnetism~\cite{Son2006, Wimmer2008, Guo2008, Wang2009, Lu2010, Ozaki2010, Saffarzadeh2011,Aprojanz2018}.

Zigzag graphene nanoribbons (ZGNRs) host spin-polarized transport channels near their edges, but similar contributions from each of their AFM-aligned edges tend to cancel any net spin signal.
Transverse electric fields~\cite{Son2006} and asymmetric edges or junctions~\cite{Wimmer2008, Sun2020, Jiang2021} are two routes to induce half-metallic behaviour and spin-filtering functionalities in these systems.
Finite dot structures allow a greater number of edges and more complex behaviour~\cite{Fernandez-Rossier2007, Bhowmick2008, Wang2009, Wimmer2010, Potasz2012}.
For example, triangular flakes can have a sublattice imbalance, resulting in a net magnetisation~\cite{Lieb1989}.
Similar behaviour is predicted for subtractive antidot systems~\cite{Yu2008b, Zheng2009, Trolle2013}, where ZZ-edged perforations underpin half-metallic~\cite{Gregersen2017} and anisotropic transport~\cite{Gregersen2018}.

\begin{figure}
\includegraphics[width =0.48\textwidth]{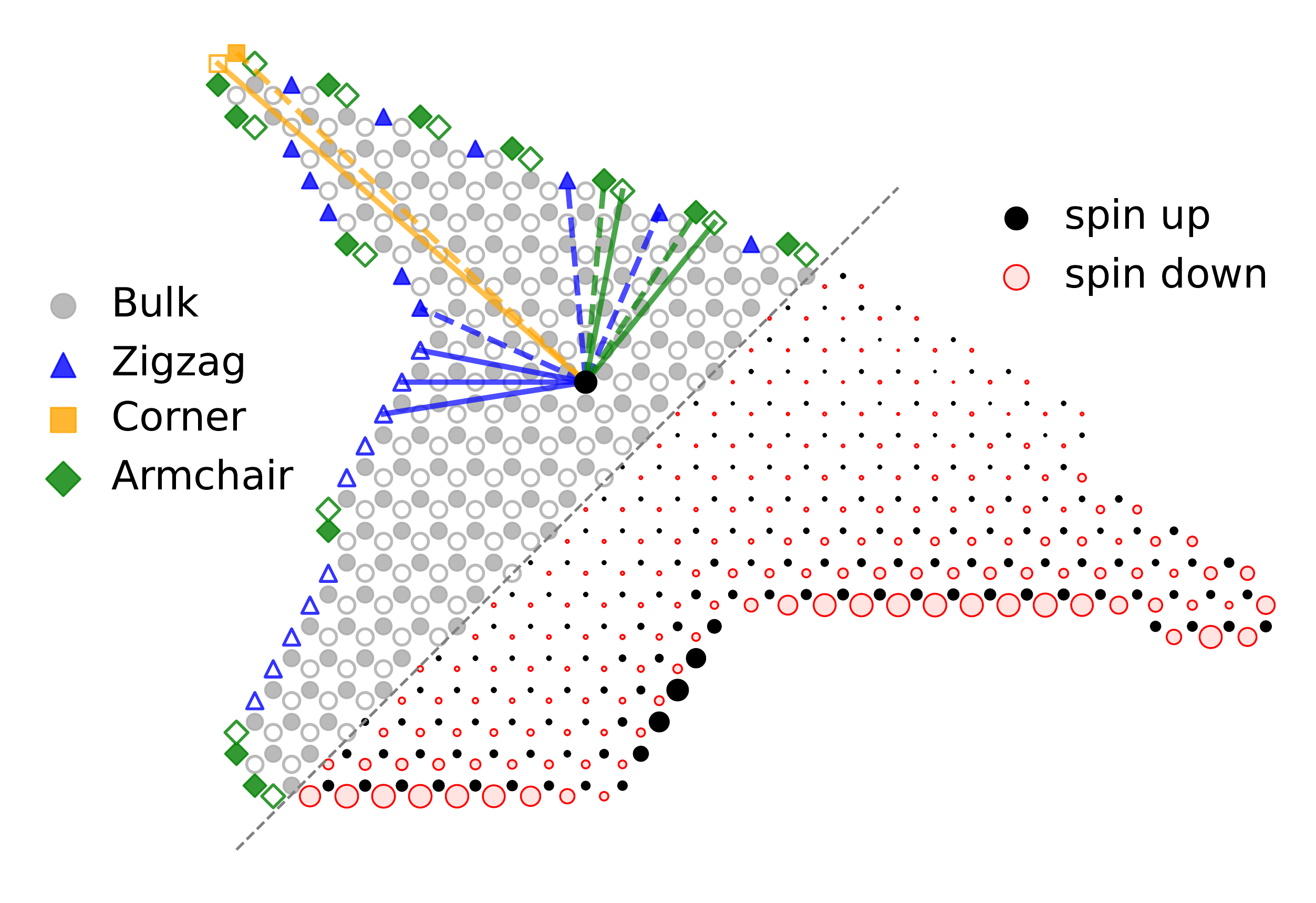}
	\caption{\textit{(Left:)} Site types within a disordered graphene flake, with sublattices shown by filled or hollow symbols. Lines show the nearest edge sites (of each type) to the black circle.
	\textit{(Right:)}
	Self-consistently calculated moments in the flake, with size (color) denoting the moment magnitude (sign).}
	\label{fig_schematic}
\end{figure}

Moment magnitudes tend to decay away from, and vary along, ZZ edges.
Maxima are found at the centre of extended ZZ sections, with smaller values towards the bulk and near junctions or corners.~\cite{Fernandez-Rossier2007, Bhowmick2008, Yazyev2011}
While fabrication methods can create edges with preferential orientations~\cite{Li2008, Kosynkin2009, Jia2009, Cai2010, Magda2014, Baringhaus2014, Ruffieux2016}, most experimental systems contain a mix of different edge lengths and types~\cite{Han2007, Tombros2011, Terres2016, Caridad2018}.
The coupling between states at these edges can lead to complex magnetic profiles, as demonstrated in Fig. \ref{fig_schematic}.

Until recently~\cite{Mishra2020}, direct experimental evidence of local magnetism had proven elusive, and most studies rely on indirect signatures from scanning tunneling measurements ~\cite{Tao2011, Magda2014,  Ruffieux2016}.
These depend sensitively on the geometry and moment profile of the flake, and experimental measurements are usually compared to theoretical calculations to verify the presence of local moments.
Moment profiles can be simulated using spin-polarised density functional theory (DFT) if the system is  very small, or can be represented by a periodic unit cell. 
Larger systems can be considered by using a tight-binding (TB) approach with a mean-field Hubbard term. 
However this scales poorly due to the repeated diagonalization of large matrices in the self-consistent (SC) procedure,  curtailing efforts to investigate large or disordered systems. 

In this work, we develop a machine-learning (ML) approach to estimate moment profiles entirely from geometric considerations.
This is motivated by the observation that moment magnitudes depend largely on their location relative to nearby edges, and to the lengths and terminating sublattices of those edges.
Similar qualitative trends are found both in both large, disordered systems and small, simple systems.
For example, the rough flake in Fig. \ref{fig_schematic} shows sublattice-dependent moments which decay towards the bulk, armchair (AC) edges and junctions, following similar trends found in pristine ribbons.
We find that neural networks, trained on simple geometric descriptors which capture these features, can accurately estimate moments in graphene flakes and ribbons, removing a troublesome computational bottleneck to calculating their electronic, magnetic and transport properties.

\paragraph{Method:}
A library of 4505 graphene flakes was created, containing both regular (finite ZGNRs) and irregular (many-sided polygons and edge-disordered dots) geometries.
This contains a variety of flake sizes (from 50 to 9000 atoms), edge lengths and edge types. 
The system in Fig. \ref{fig_schematic} is produced by the polygon method, with examples of the other types given in \cite{suppmat}.
Each site in a flake is classified as either `Bulk' or one of three edge types: `Zigzag', `Corner', or `Armchair'~\cite{Singh2011}.
A nearest neighbour TB model is used to describe the electronic structure of the flakes, with an onsite Hubbard term capturing electron-electron interactions.
Within the mean-field approximation, this reduces to a spin-dependent onsite potential $\epsilon_{i\sigma} = - \sigma \tfrac{U}{2} m_i$ at each site $i$, where $m_i$ is the local moment and a Hubbard parameter $U=1.33 \lvert t \rvert$ gives results in good agreement with DFT calculations~\cite{Yazyev2010}. 
In this work, we consider the energetically favourable AFM solution and restrict our focus to the undoped case with half-filled $p_z$ orbitals. 
The values of $m_i$ are calculated by a self-consistent (SC) procedure~\cite{suppmat}, starting from an initial trial solution.

The SC calculation creates a mapping from a flake geometry, via the Hamiltonian, to a moment profile.
Deep neural networks can be a powerful tool to approximate such mappings where their exact form is unknown or prohibitively expensive to calculate.
A network consisting of layers of artificial neurons is exposed to a training data, and has its parameters adjusted to minimise the error between the output and the expected result.
A network is tested on unseen data during (`validation') and after (`test') the training process to prevent over-fitting and establish its accuracy.
Such methods have been used to estimate ab-initio figures-of-merit for materials design~\cite{Hansen2013, Schmidt2019}, reproduce the mapping performed by functionals in complex many-body models~\cite{Nelson2019}, and predict critical behaviour in lattice models~\cite{Carrasquilla2017, Deng2017}.
In graphene systems, ML approaches have been employed to directly predict transport properties in the presence of disorder~\cite{Lopez-Bezanilla2014}, strain~\cite{Torres2019, Nedell2022} or doping~\cite{Ryczko2020}.

\begin{figure}
	\includegraphics[width =0.49\textwidth]{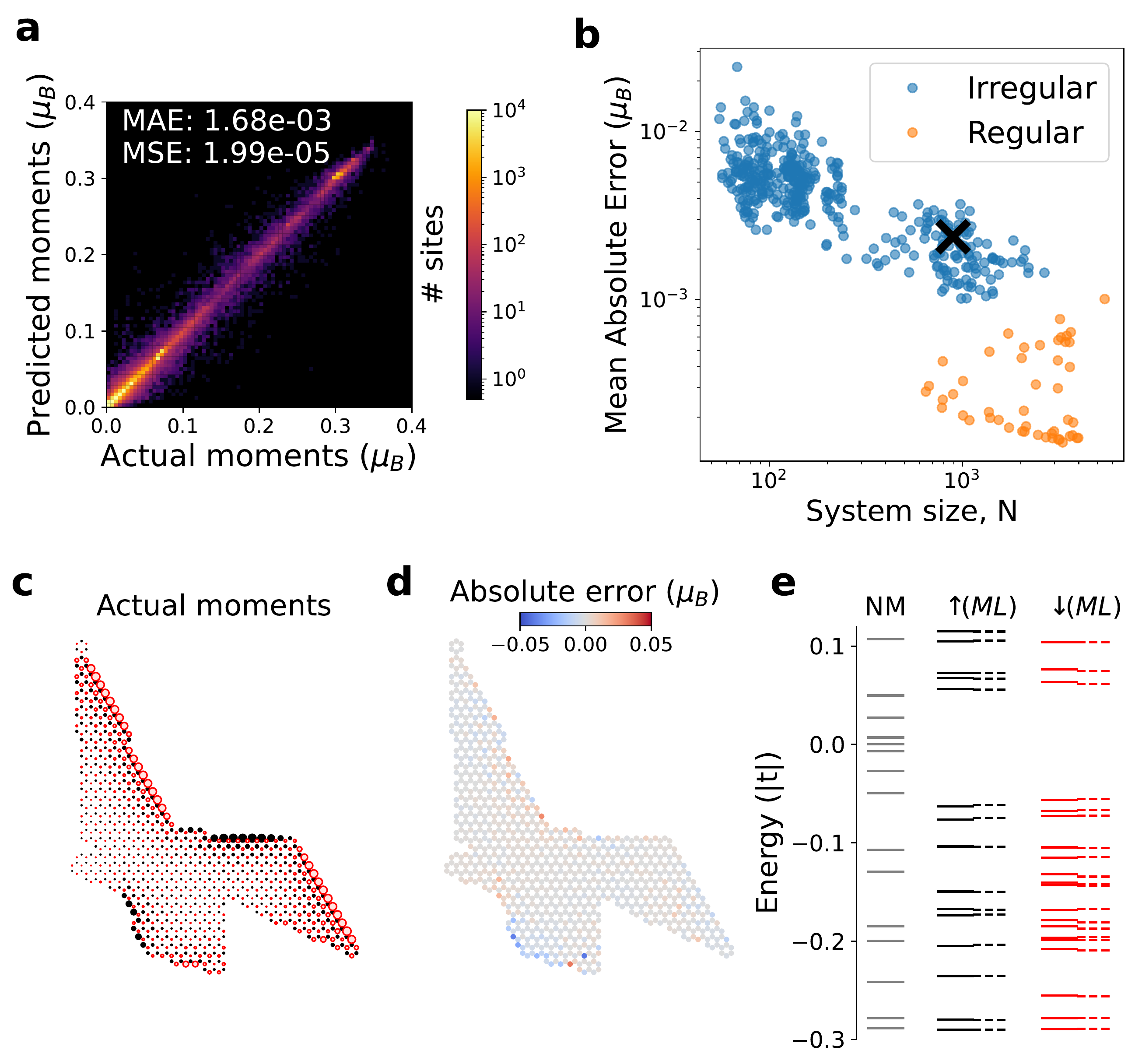}
	\caption{\textbf{(a)} Comparison of actual and NN-predicted moment magnitudes for every site, in every flake, in the test set. \textbf{(b)} Mean absolute error (MAE) of every system in the test set, according to flake size and etching method. \textbf{(c,d)} Maps of the actual moments and the deviation of the predicted moments for a test system. \textbf{(e)} Comparison of spin-polarized energy levels of this system calculated using the actual (solid lines) or predicted moment profiles  (dashed lines). Grey lines show the energy levels for a non-magnetic calculation.}
	\label{fig_results}
\end{figure}

Since the magnetic moments in a flake are largely influenced by local geometric details, we train neural networks~\cite{tensorflow2015-whitepaper, chollet2015keras} to return individual site moments, based on local descriptors, rather than the moment profiles of entire structures.
This allows trained models to be easily applied to systems of arbitrary size.
The descriptor for a site includes the distances to nearby edge sites as shown in Fig. \ref{fig_schematic}, and between these edge sites.
The number of bonds along the shortest path between two sites is used as the distance metric, capturing the connectivity of the lattice, which is important for more complex geometries.
The descriptor also takes sublattice indices into account and is invariant under rotations and translations.
The network is trained to identify the absolute value of a moment, with its sign determined by sublattice.
See \cite{suppmat} for more details on descriptor generation, and an online code to make moment predictions for arbitrary geometries.

\begin{figure*}
	\includegraphics[width =0.98\textwidth]{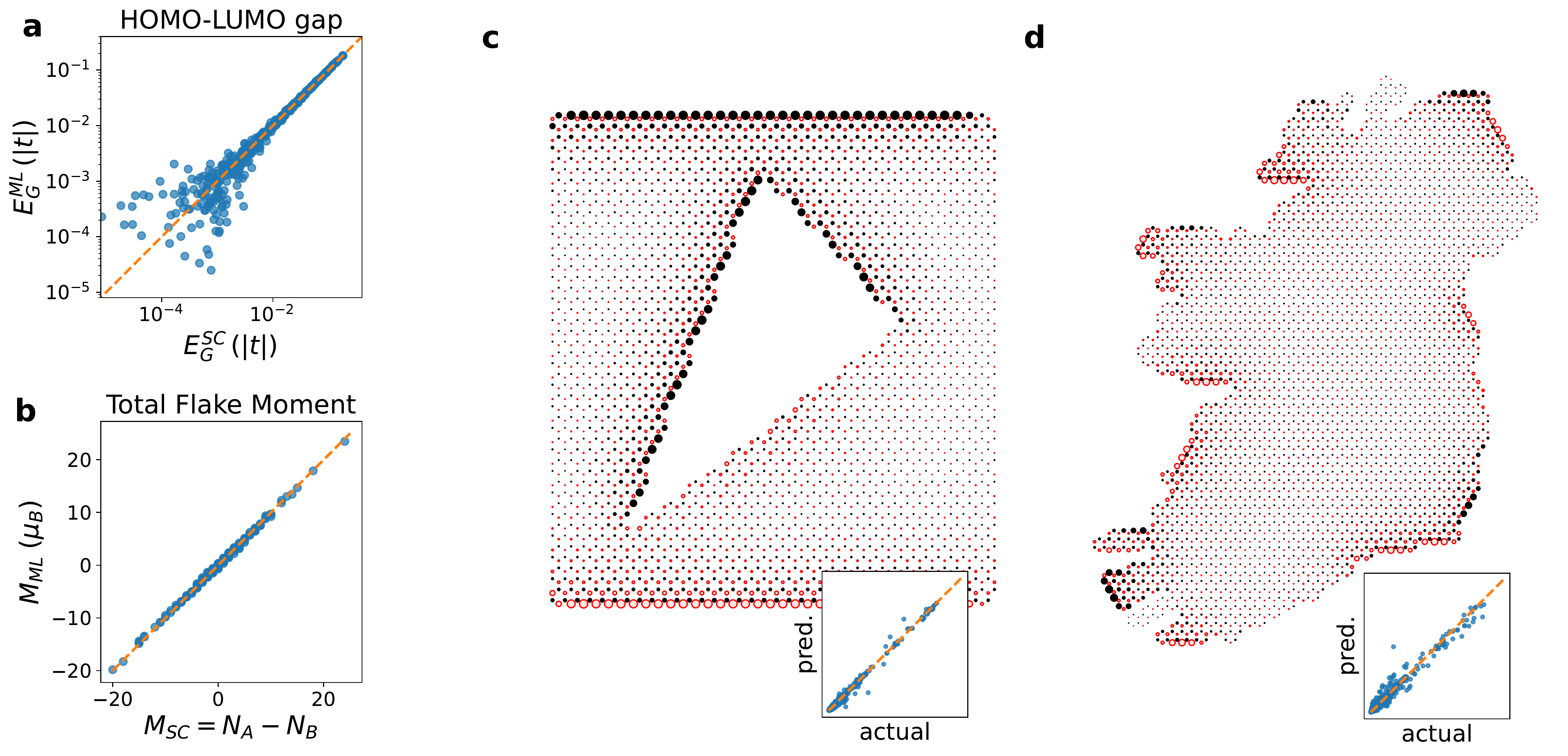}
	\caption{\textbf{(a,b)} Comparison of the spin-split band gap and the total magnetic moment of each flake in the test set, calculated using moments from both self-consistent (SC) and machine-learning (ML) methods. \textbf{(c,d)} Predicted moments for geometries with features, such as internal edges and rugged peninsulas, which are not present in the training set. The insets compare the actual and predicted moments in each system.}
	\label{fig_generalization}
\end{figure*}

The actual moments for all 271,590 sites in the 451 test set flakes are compared to those predicted by a trained neural network in Fig. \ref{fig_results}a.
The distribution is maximal along the diagonal and decays rapidly away from it, showing an excellent correspondence between predicted and actual moments.
While the mean absolute error (MAE) is a useful metric to compare networks trained on the same set, its magnitude largely depends on the specific flakes in the test set.
The MAE for a single flake depends on its size (Fig. \ref{fig_results}b), as larger systems have, proportionally, more edge sites with larger moments and prediction errors.
The MAE decays as $\sim 1/\sqrt{N}$ for irregular flakes, where $N$ is the number of sites in the flake, confirming that the MAE follows the relative edge length of the flake.
Regular geometries have a much smaller MAE due to their uniform moment profiles.
A more meaningful metric of the network's accuracy is the MAE on all edge sites in irregular flakes, which is found to be $\sim 7 \times 10^{-3}\mu_B$, or less than 3\% of the edge moment in a pristine ZGNR~\cite{suppmat}.

\paragraph{Generalization:}
The actual moments in a polygon flake from the test set and the prediction error for each moment are mapped in Fig. \ref{fig_results}c,d.
This system corresponds to the black `X' in Fig. \ref{fig_results}b. 
The predicted moments display the expected trends, \emph{i.e.} decay from the centre of long zigzag edges towards the bulk or other edge types.
The quantitative match with the actual moments is excellent, with discrepancies at only very few sites.
Since the network is only given local information about each site, it is not aware of the overall size, shape or edge length of a flake.
However, many quantities depend on the moment profile of the entire flake, and if sufficiently accurate, the  more-or-less instantaneous ML predictions would allow us to circumvent the SC computational bottlenecks.
Energy spectra of the considered flake are shown in Fig. \ref{fig_results}e, with the dashed black and red lines showing the up- and down-spin levels calculated using ML moments.
They align almost perfectly with results using exact SC moment profiles (solid lines).
Comparison with the non-magnetic case (grey lines) shows that the ML levels accurately capture both the qualitative and quantitative features of the spectrum, including the spin-splitting of zero-energy states and the overall level distribution.

The band gap $E_G$ of a flake can be enhanced by spin-splitting, and is sometimes taken as a proxy for local magnetism in experiments~\cite{Magda2014}.
Fig. \ref{fig_generalization}a compares $E_G$ (in log scale), calculated using each set of moments, for every test set flake.
The ML moments give a highly accuracy estimate once $E_G \gtrsim 0.01 |t| \sim 30 $meV, with an average error of $\sim 3 \%$.
The total magnetic moment $M$ of a flake is also of interest, since the SC value exactly obeys Lieb's theorem~\cite{Lieb1989} and is given by the difference between the number of $A$ and $B$ sublattice sites: $M= N_A - N_B$.
Site-by-site ML estimates place no constraints on $M$, nonetheless Fig. \ref{fig_generalization}b shows excellent agreement with SC results and Lieb's theorem.

The ML approach clearly allows for quick and reliable estimations of system-wide properties relevant to both theoretical and experimental studies.
The model successfully generalises to flakes similar to those on which it was trained, but as these do not include every possible kind of edge profile, it is worth examining its performance outside the test set.
Fig. \ref{fig_generalization}c contains an internal perforation, none of which are present in the training set, and which increases the complexity of the moment profile.
Nonetheless the network is able to accurately reproduce the SC result.
Key to its success is the distance metric used in the descriptor, which prevents undue influence between sites separated by the perforation.
This is also important for the geometry in Fig. \ref{fig_generalization}d, based on the coastline of Ireland.
This contains a mix of edges, angles and local disorders that do not emerge from the simpler etching methods.
ML predictions again agree with full calculations, and in particular capture complex behaviour where parallel peninsulas give large numbers of nearby edge sites. 

\paragraph{Application to spin transport:}
Finite flakes are a useful platform for building a ML model, but applicability is limited without generalisation to extended ribbons, which underpin the majority of proposed devices. 
An open question is whether spin-polarised transport will survive in realistic systems.
Local defects, functionalizations and reconstructions can inhibit moment formation or backscatter electrons near the edge, removing the desired behaviour~\cite{Wimmer2008, Huang2008, Kunstmann2011, Pizzochero2021}.
State-of-the-art epitaxial~\cite{Baringhaus2014, Aprojanz2018} and bottom-up grown ribbons~\cite{Li2008, Cai2010, Ruffieux2016}, where such strong local disorders are largely absent, nonetheless include longer-ranged disorders, like smooth width fluctuations or irregular junctions. 
Edge roughness has been extensively studied in non-magnetic ribbons and typically requires  configurational averaging over large numbers of ZGNRs~\cite{Mucciolo2009}.
This inhibits similar studies in magnetic ribbons, as calculating the required moment profiles is prohibitively expensive.
This cost could be allayed using accurate ML moments, enabling in-depth study of spin-polarised transport in disordered ZGNRs. 

\begin{figure}
	\includegraphics[width =0.48\textwidth]{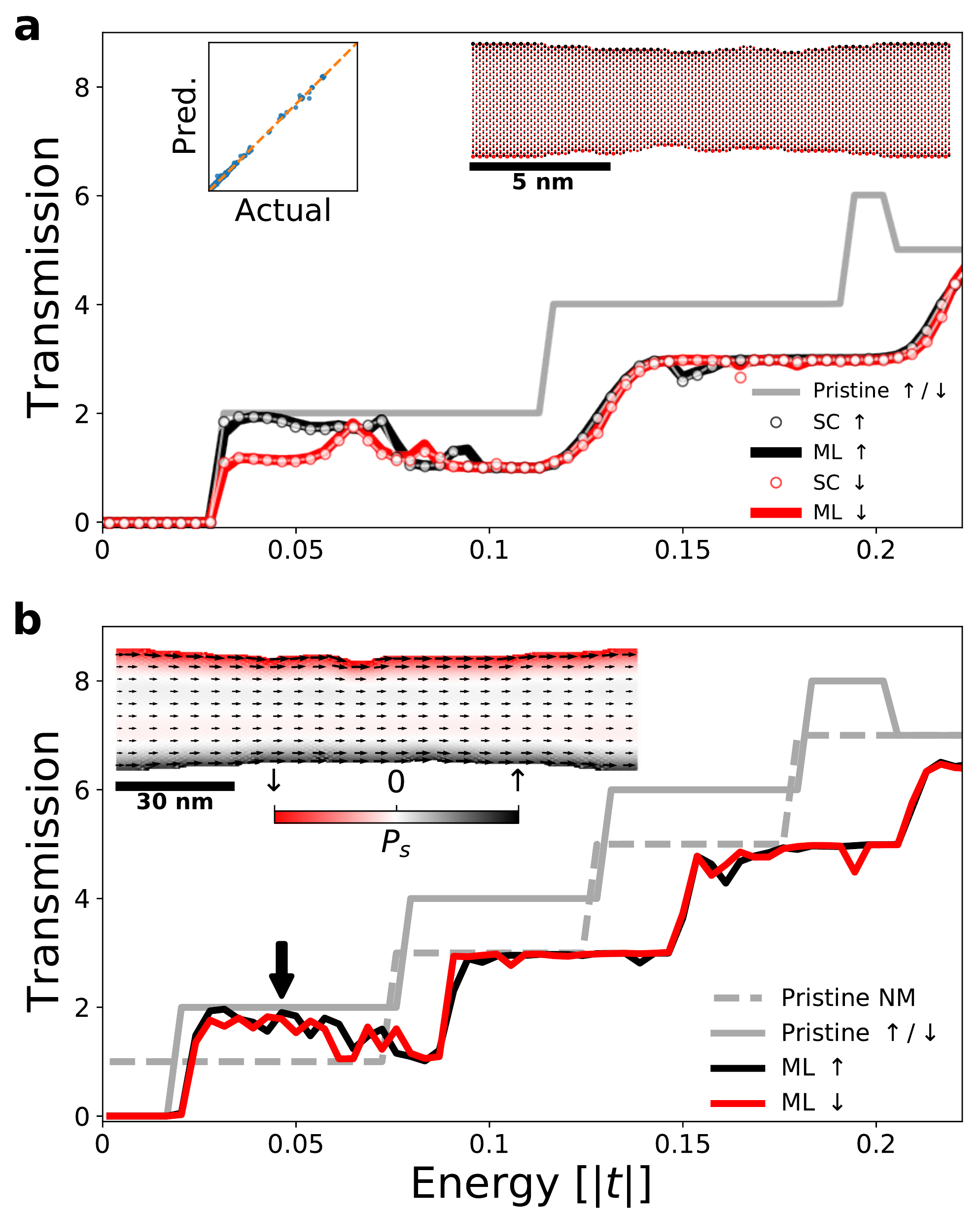}
	\caption{\textbf{(a)} Spin-dependent transmission in an edge-disordered ZGNR, calculated using SC (symbols) and ML (lines) moment profiles. The insets show the accuracy of the ML moments and the moment profile along the ribbon. 
	\textbf{(b)} Transmissions through a larger system. The inset maps the current flow and spin polarisation at the marked energy. Spin-polarised channels are visible along the rough edges.
	}
	\label{fig_transport}
\end{figure}

The device in Fig. \ref{fig_transport}a contains smooth width fluctuations, similar to that observed in experimental systems~\cite{Caridad2018, Aprojanz2018}.
It is also near the upper limit of systems that can be solved self-consistently, allowing us to benchmark ML results.
The inset confirms the the excellent generalisation of the ML model to this system, which manifests as an almost perfect agreement between transmission calculations using ML (lines) and SC (dots) moments.
Note that we are only using ML to predict moment profiles, not to directly estimate transmission~\cite{Lopez-Bezanilla2014}, as this can be efficiently calculated using recursive techniques~\cite{Papior2017, suppmat}.
Since the ML moments are sufficiently accurate for transport calculations, we can now scale up to larger systems that would otherwise be computationally intractable.
Fig. \ref{fig_transport}b investigates such a system using ML-predicted moments only.
Without edge disorder, the transmission (per spin) is higher in a magnetic (solid grey) than a non-magnetic ribbon (dashed grey) in the range $ 0.02 \lvert t \rvert \lesssim  E \lesssim 0.2\lvert t \rvert$.
This is due to an additional edge-localised dispersive band, arising from spin-splitting of a flat band in the non-magnetic system.
An additional mode at this energy, for both ribbons, carries current of both spin-orientations at the ribbon centre.

Comparing the disordered spin-dependent results to both pristine ribbons allows us to explain the effect of edge roughness.
At energies $E\gtrsim 0.1\lvert t \rvert$, the disordered transmissions return to the sequence seen for NM ribbons, suggesting that the spin-polarised edge mode has been completely quenched.
The other modes are robust against smooth disorder, so the plateau behaviour persists, but with an energy shift caused by mode-matching between sections with different widths.
Very different behaviour is seen at low energies, \emph{e.g.} at the arrow in Fig. \ref{fig_transport}b, where the disordered spin-dependent cases show near perfect transmission.
The persistence of spin transport channels here is due to a lack of backscattering possibilities.
The left- and right-propagating modes for a given spin are located on the same edge, so scattering between these modes can occur without a spin flip.
However, these two modes are associated with widely-separated states in reciprocal space, so that scattering between them requires short-ranged disorder, which is not present here.
The other mode at this energy has very little edge component, which limits backscattering into it.

This suppression of backscattering allows spin-polarized transport in edge channels to persist, as shown explicitly in the inset map, which plots the local current flow (arrows) and spin polarization (colour) for the marked energy.
While spin transport survives at low energies, it vanishes at higher energies due to the onset of higher-order bulk modes which offer additional backscattering possibilities.
These results provide a useful insight for experimental device design: the desired spintronic behaviour should appear also for rough edges, but in a more limited energy range than for pristine systems. 

\paragraph{Conclusions:}
Neural networks gives a quick and reliable estimate of the mean-field solution to the Hubbard model in a half-filled bipartite lattice.
Prediction cost scales linearly with system size, allowing for rapid calculations in systems that were previously intractable.
Immediate applications lie in the study of graphene edge magnetism, where the electronic, magnetic and transport properties of a wide range of systems can be efficiently and accurately calculated.
Given the excellent generalization of the model, both to geometries and quantities unseen during training, it can be applied to a number of problems.
It allows for a detailed analysis of transport in disordered GNRs, where understanding the interplay between large-scale disorder and local magnetism is essential if GNRs are to play their mooted roles in electronic, spintronic and quantum devices.
To help achieve this, we have made available sample code to quickly apply our model and predict moments for arbitrary geometries~\cite{suppmat}.   

Our approach can also act as a starting point for more accurate studies or more complex problems.
ML profiles can be used as almost-converged `initial guesses' in SC calculations, reducing the number of steps required.
More accurate ML estimates may also be achieved by supplementing the simple geometric information in our descriptors.
For example, moment formation is strongly tied to the local density of states~\cite{Anderson1961}, so including a non-SC evaluation of this quantity could improve model performance.
A trade-off will emerge here between the extra time needed to generate the descriptor and the accuracy required.
These strategies were not needed to achieve high accuracy in this work, but could help in more complicated cases.
Examples include lattices away from half-filling, with non-uniform background potentials, or of different materials, where exact solutions are difficult to calculate and display less intuitive trends.

Finally, we focused here on material properties at the nano- and mesoscopic scales where graphene systems are investigated experimentally and incorporated into devices.
This is in contrast to previous studies using ML to consider more fundamental aspects of the Hubbard model~\cite{Saito2017, Nelson2019}, which typically consider smaller systems in which exact solutions can be found.
However, the techniques developed here are not scale specific and may also find applicability in advanced many-body models.

\begin{acknowledgments}
The authors wish to acknowledge funding from the Irish Research Council under the Laureate awards programme.
\end{acknowledgments}


%

\end{document}